\documentclass[11pt]{article}

\usepackage{SIGACTNews}

\usepackage{chicagor}
\usepackage{amsmath}
\usepackage{amssymb}

\def\IssueDate{December 2004}
\def\IssueVolume{35}
\def\IssueNumber{4}

\newcommand{\thmcolon}{}

\newtheorem{THEOREM}{Theorem}%
\newtheorem{LEMMA}[THEOREM]{Lemma}
\newtheorem{COROLLARY}[THEOREM]{Corollary}
\newtheorem{PROPOSITION}[THEOREM]{Proposition}
\newtheorem{DEFINITION}[THEOREM]{Definition}
\newtheorem{CLAIM}[THEOREM]{Claim}
\newtheorem{EXAMPLE}[THEOREM]{Example}
\newtheorem{REMARK}[THEOREM]{Remark}

\newenvironment{theorem}{
  \begin{THEOREM}\hspace{-1ex}{\bf .} }%
                        {\end{THEOREM}}
                      {\end{LEMMA}}
                          {\end{COROLLARY}}
                            {\end{PROPOSITION}}
                            { \end{DEFINITION}}
                            {\end{CLAIM}}

                            { \wbox\end{EXAMPLE}}
\newenvironment{example*}{\begin{EXAMPLE} \thmcolon  \rm}%
                            {\end{EXAMPLE}}
                            { \wbox\end{REMARK}}
\newenvironment{remark*}{\begin{REMARK} \thmcolon  \rm}%
                            {\end{REMARK}}

\def\squareforqed{\hbox{\rlap{$\sqcap$}$\sqcup$}}
\def\wbox{\ifmmode\squareforqed\else{\unskip\nobreak\hfil
\penalty50\hskip1em\null\nobreak\hfil\squareforqed
\parfillskip=0pt\finalhyphendemerits=0\endgraf}\fi}

\newcommand{\Rule}[2]{          %
  \begin{array}{c}
  #1 \\\hline
  #2
  \end{array}}

\newcommand{\derives}{\vdash}
\newcommand{\nderives}{\nvdash}
\newcommand{\encr}[2]{\{#1\}_{#2}} 
\newcommand{\inv}[1]{#1^{-1}}

\usepackage{url}

\newcommand{\sat}{\models}
\newcommand{\rimp}{\Rightarrow}

\renewcommand{\phi}{\varphi}

\renewcommand{\emptyset}{\varnothing}
\renewcommand{\Pr}{\mathrm{Pr}}

\newcommand{\cK}{\mathcal{K}}
\newcommand{\cM}{\mathcal{M}}

\newcommand{\adv}{{\scriptscriptstyle\mathit{adv}}}
\newcommand{\local}{{\scriptscriptstyle\mathit{local}}}
\newcommand{\crypt}{{\scriptscriptstyle\mathit{crypt}}}
\newcommand{\dy}{{\scriptscriptstyle\mathit{dy}}}

\title{SIGACT News Logic Column 10\\[2ex]
\textbf{Specifying Confidentiality}\footnote{\copyright{} Riccardo
Pucella, 2004.}} 
\author{Riccardo Pucella\\
Cornell University\\
Ithaca, NY 14853 USA\\
riccardo@cs.cornell.edu}
\date{}

\begin{document}

\SIGACTmaketitle

I would like to start my tenure as editor of the Logic Column by
thanking Jon Riecke, who has edited this column since 1998. The Logic
Column serves as a showcase of the many connections between logic and
computer science. Logic has been connected with computer science since
the early days of Turing. In the past few decades, logical methods
have had a considerable impact. To get a sense of the range of
applications, consider the 2001 NSF/CISE Workshop on The Unusual
Effectiveness of Logic in Computer Science (see
\url{http://www.cs.rice.edu/~vardi/logic/}). An article derived from
the workshop appeared in the \emph{Bulletin of Symbolic Logic}
\shortcite{r:halpern01d}, and it is an exceedingly good read. If you get a
copy of that issue of the Bulletin, make sure to also have a look at
the article by Buss et al. \citeyear{r:buss01}, which discusses the
current state of mathematical logic. 

If you have any suggestion concerning the content of the Logic Column,
or even better, if you would like to contribute by writing a
survey or tutorial on your own work or topic related to your area of
interest, feel free to get in touch with me. Topic of interest
include, but are not limited to: 

\begin{itemize} 

\item recent results on logic in general, and in applications to 
computer science in particular;

\item reviews of research monographs and edited volumes; 

\item conference reports;

\item relevant results and connections with other fields that make use
of logical methods, such as mathematics, artificial intelligence,
linguistics, and philosophy; 

\item surveys of interesting uses of logical methods in computer
science. 

\end{itemize}

And while we are on the topic of logical methods in computer science,
let me take this opportunity to advertise a new online journal, aptly
called \emph{Logical Methods in Computer Science}. See
\url{http://www.lmcs-online.org/} for more details.

\section*{Modeling Confidentiality}

First-time novelists write transparently autobiographical novels;
first-time columnists write about what they do. Therefore, this
article will be about logical methods applied to security, a topic I
have been involved with for the past few years. The goal is to
illustrate a feature of logical methods: to unify under the umbrella
of a formal language seemingly distinct notions that share a common
intuition. Some of the results I report reflect ongoing work in
collaboration with Sabina Petride, of Cornell University.

There are a number of important concepts in security: confidentiality
(keeping data secret), authentication (proving identity or
origination), integrity (preventing data modification), and others. In 
this article, I will focus on confidentiality, arguably one of the core
concepts. 
There are many views on confidentiality in the literature, with many
corresponding definitions, in many different guises. My goal here is
to show that these definitions can essentially be understood as follow:
they capture the fact that unauthorized agents do not \emph{know}
anything about confidential data. The variations between
definitions concern the kind of data being protected, and the
properties of the data that are meant to remain unknown. The
setting where the various definitions will be interpreted is a setting
where we can make sense of such knowledge, in a pleasantly abstract
way.

The first step is to specify what we mean by an unauthorized
agent. Generally, security is studied in an \emph{adversarial}
setting, that is, in the presence of an adversary. Given our focus on
confidentiality, we assume that the adversary seeks to circumvent
confidentiality, and obtain information about the confidential 
data. To simplify our problem slightly, we assume that
confidentiality is meant to be enforced against such an adversary, and
thus that ensuring confidentiality means that the
adversary does not know anything about a confidential piece of data.

The second step is therefore to capture the knowledge of the adversary
in some general way.  A particularly successful formalization of
knowledge is due to Hintikka
\citeyear{r:hintikka62}, and has been applied to many fields of
computer science; see Fagin et al. \citeyear{r:fagin95} for a
survey. The formalization relies on the notion of \emph{possible
worlds}: a possible world is, roughly speaking, a possible way in
which the world could be. To drive the intuition, consider the
following situation. Suppose I witness Alice murdering Bob in the
library, and suppose that it is a matter of fact that Alice used a
fire poker, but for whatever reason, I did not notice the murder
weapon that Alice used. Thus, there are (at least) two worlds that I
consider as possible alternatives to the actual world: the actual
world itself, where Alice used a fire poker, and a world where Alice
used, say, a brick.  I cannot be said to know that Alice murdered Bob
using a poker, since from my point of view, it is possible that Alice
did not: there is a world I consider possible where Alice used a
brick. On the other hand, I can be said to know that Alice is a
murderer, since it will be the case at every world I consider
possible.\footnote{This is under the assumption that I am not subject
to illusions, or hallucinations, of course. Philosophers are fond of
such counterexamples, which reveal implicit assumptions about the
world that may affect our reasoning. When applying these ideas to
computer science, we shall assume that our models take into account
everything relevant to establish knowledge, including such implicit
assumptions.}  Thus, I can be said to \emph{know} a fact at a world if
that fact is true in all the worlds I consider possible at that
world. 

To reason about possible worlds and the knowledge of an agent with
respect to these worlds, we use epistemic frames. An \emph{epistemic
frame} is a tuple $F=(W,\cK)$, where $W$ is a set of possible worlds
(or states) and $\cK$ is a relation on $W$ that represents the worlds
that the agent considers as possible alternatives to other worlds;
$(w,w')\in\cK$ if the agent consider $w'$ as a possible world at world
$w$.  We often use the notation $\cK(w)$ for $\{w' \mid
(w,w')\in\cK\}$. 
We identify a fact with the set of
worlds where that fact holds. Thus, a fact is a subset $S$ of
$W$. Following the intuition above, we say a fact $S$ is \emph{known}
at a world $w$ if $S\subseteq\cK(w)$, that is, if at every world that
the agent considers possible at world $w$, the fact $S$ holds at that
world.  To model the situation in the previous paragraph, consider a
simple epistemic frame with three worlds $W=\{w_1,w_2,w_3\}$, where
$w_1$ is the world where Alice murdered Bob in the library with a
poker, $w_2$ is the world where Alice murdered Bob in the library with
a brick, and $w_3$ is the world where Alice did not murder Bob. Thus,
the fact $G_1$=``Alice murdered Bob in the library'' is represented by
the set $\{w_1,w_2\}$, and the fact $G_2$=``Alice murdered Bob with a
poker'' is represented by the set $\{w_1\}$. By assumption, the worlds
I consider possible at $w_1$ are $\{w_1,w_2\}$, and thus
$\cK(w_1)=\{w_1,w_2\}$. Since $\cK(w_1)=\{w_1,w_2\}\subseteq G_1$, I
know the fact $G_1$, but since
$\cK(w_1)=\{w_1,w_2\}\nsubseteq\{w_1\}$, I do not know $G_2$.

The framework can be easily extended to reason about the knowledge of
multiple agents. It suffices to provide a relation $\cK_i$ to every
agent $i$. In this article, since we shall focus on confidentiality
with respect to only a single adversary, we only need to reason about the
adversary's knowledge. This has the advantage of simplifying the
framework and the notation. Of course, our discussion can be expanded
to deal with multiple agents. In fact, we will assume multiple agents,
but only model the knowledge of the adversary.

Epistemic frames describe the structure of the model that we want to
reason about. They have been quite successful in fields such as
economics, where they are used to reason about the knowledge of
economic agents \cite{r:aumann99}.  While a lot can be done purely at
the level of the model, one big advantage in casting a situation in
epistemic frames is that we can define a formal language to let us do
the reasoning without having to explicitly manipulate the worlds of
the model. The language of \emph{epistemic logic} starts with a set of
primitive propositions $\Phi_0$ (describing the basic facts we are
interested in reasoning about), and forming more general formulas
using conjunction $\phi\land\psi$, negation $\neg\phi$, and knowledge
formulas of the form $K\phi$, read ``the agent knows
$\phi$''. In order to interpret this language in an epistemic frame,
that is, to say when a formula of the language is true at a world of the
frame, we need to add an \emph{interpretation} $\pi$ stating which
primitive propositions are true at which worlds. An \emph{epistemic
structure} (also known as a Kripke structure) is a tuple
$M=(W,\cK,\pi)$, where $(W,\cK)$ is an epistemic frame, and $\pi$ is
an interpretation. The truth of a formula $\phi$ at a world $w$ of
structure $M$, written $(M,w)\sat\phi$, is established by induction on
the structure of $\phi$:
\begin{itemize}
\item[] $(M,w)\sat p$ if $p\in \pi(w)$
\item[] $(M,w)\sat \neg\phi$ if $(M,w)\not\sat\phi$
\item[] $(M,w)\sat \phi\land\psi$ if $(M,w)\sat\phi$ and $(M,w)\sat\psi$
\item[] $(M,w)\sat K \phi$ if for all $w'\in\cK(w)$, $(M,w')\sat\phi$.
\end{itemize}
The semantics for primitive propositions shows the role of the
interpretation.  The semantics of negation and conjunction are the
obvious ones. The semantics of knowledge formulas follows the
intuition outlined above: a formula is known at a world $w$ if it is
true at all the worlds the agent considers possible at $w$. We write
$M\sat\phi$ if $(M,w)\sat\phi$ for all worlds $w$. 

We therefore have two tools to reason about the knowledge of agents: a
way to model the system with a notion of possible worlds, and a
language  to express properties of the
system. These two tools come together when applying the framework to
capture various notions of confidentiality in the literature.

Rather than using epistemic structures in their full generality, we
focus on a particular class of structures, inspired by the multiagent
systems often used to model distributed systems.
We assume a number of agents (named $1$ to $n$, for simplicity),
including an adversary, named $\mathit{adv}$.  We assume that every
agent (including the adversary) is in some local state at any global
state of the system. We take as the worlds of our model the global
states of the system. We furthermore assume that the environment acts
like an agent, and has its own local state, to account for the
information that needs to be maintained but is not kept in the local
state of any agent. Thus, a global state is a tuple
$(s_e,s_\adv,s_1,\dots,s_n)$, where $s_e$ is the local state of the
environment, $s_\adv$ is the local state of the adversary, and $s_i$
is the local state of agent $i\in\{1,\dots,n\}$. Intuitively, the
local state of an agent represents the part of the global state that
he can observe. Thus, if an adversary has the same local state in two
global states $s,s'$, then at state $s$, he should consider state $s'$
as a possible global state, since he can observe exactly the same
local state in both cases. In other words, the basic possible worlds
relation for the adversary (called an
\emph{indistinguishability relation} since it is based on the idea of
distinguishing local states) we consider is the state-identity
relation $\cK^\local$, which holds between two states if the adversary
has the same local state in both states. Formally,
$(s,s')\in\cK^\local$ if only only if $s_\adv=s'_\adv$.  While we will
find it useful to customize the indistinguishability relation of the
adversary to control what he can observe, especially when dealing with
cryptography, we shall always assume that the adversary cannot
distinguish two states where he has the same local state. Thus, if
$\cK$ is an indistinguishability relation for the adversary, we have
$(s,s')\in\cK$ whenever $s_\adv=s'_\adv$, that is,
$\cK^\local\subseteq\cK$. Putting this all together, we define an
\emph{adversarial frame} as a tuple $F=(S,\cK)$, where $S$ is a set of
global states, and $\cK$ is an indistinguishability relation for the
adversary with $\cK^\local\subseteq\cK$.  Similarly, an
\emph{adversarial structure} is a tuple $M=(S,\cK,\pi)$ where
$(S,\cK)$ is an adversarial frame, and $\pi$ is an
interpretation. Since adversarial structures are just epistemic
structures, we can interpret an epistemic language over adversarial
structures, where the knowledge operator captures the knowledge of the
adversary.

We now explore how we can use this framework to explicate many
definitions of confidentiality used in the security literature. As we
shall see, all the definitions will be captured semantically, that is,
by describing appropriate conditions on adversarial structures, as well
as describing appropriate indistinguishability relations for the
adversary. Moreover, we will give an interpretation to these semantic
conditions in terms of formulas of an epistemic logic. Roughly
speaking, this interpretation means the adversary never knows a
particular class of formulas; these formulas represent properties of
the data defined to be confidential. 

\section*{Confidentiality and Information Flow}

A particular form of security is to ensure the confidentiality of
information among users at different security levels. (This is often
called \emph{multilevel security}.) An example is the stereotypical
classification of users (and data) in military systems, where security
levels include ``unclassified'', ``classified'', and ``top-secret'';
users at a given level can access information marked at that level,
and at lower levels.
The model of the world is that these users share the same system, and
the goal is to prevent the system from leaking information about the
high-level secrets to lower levels. Generally, these security levels
form a hierarchy \cite{r:denning76}. Consider the following example. A
company operates a large computer network. Alice, the CEO, has access
to all the data in the company. Bob, a consultant, uses the same
system, but has restricted access. The company would like to ensure
that Bob cannot gain any information about some of the high-level
data that Alice enters in the system. That is, the company would like
to prevent any sort of flow of information from high-level data to
low-level users.  In this setting, a confidentiality property
specifies which flows of information are allowed, and which are
forbidden.
The most general form of confidentiality is
to forbid any kind of information to flow from the high-level
users to lower-level users. In the discussion that follows, I will
suppose that there are only two classes of security levels, \emph{high} and
\emph{low}, with the adversary being a low user; however, the ideas
readily generalize to multiple security levels.

The general approach goes back to the notion of nondeducibility
introduced by Sutherland \citeyear{r:sutherland86}.  Halpern and
O'Neill \citeyear{r:halpern02a} have formalized this notion (and
others) using possible worlds and epistemic logic. I describe one of
their results in this section.
To a first approximation, the intuition is that  the low
agent should not be able to rule out possibilities as far as the
interesting part of the local state of the high agents is
concerned. To capture the interesting part of the local state of the
high agents, define an \emph{information function} for agent $i$ to be
a function $f$ on global states that depends only on agent $i$'s local 
state, that is, $f(s)=f(s')$ if $s_i=s'_i$. 

If an information function for agent $i$ describes the interesting
aspects of the local state of agent $i$ that he seeks to keep
confidential, then we can define $f$-secrecy with respect to the
adversary:\footnote{Strictly speaking, $f$-secrecy is defined with
respect to any agent, but we already established that we care only
about the adversary in this article.} agent $i$ \emph{maintains
$f$-secrecy} in an adversarial frame $F$ if for all global states $s$
and all values $v$ in the image of $f$,
\[ \cK(s)\cap f^{-1}(v)\ne\emptyset.\]
In other words, 
the adversary considers all values of $f$ possible, at every global
state.

This fairly simple semantic characterization can be captured
syntactically, using the epistemic logic described earlier, in a way
that relates to the adversary's knowledge about the high-level state.
Let $\Phi_0$ be an arbitrary set of primitive propositions. If $f$ is
an information function for agent $i$, a formula $\phi$ in $M$ is said
to be \emph{$f$-local} if it depends only on the value of $f$, that is,
whenever $f(s)=f(s')$, then $(M,s)\sat \phi$ if and only if
$(M,s')\sat\phi$. Thus, in some sense, $\phi$ is a proposition that
captures a property of the value of $f$.  Of course, we have to
account for the possibility that $\phi$ is completely trivial. Say
$\phi$ is \emph{nontrivial} in $M$ if there exists $s,s'$ such that
$(M,s)\sat\phi$ and $(M,s')\sat\neg\phi$.  The following result is
proved by Halpern and O'Neill \citeyear{r:halpern02a}.

\begin{theorem}\label{t:kevin} Let $F=(S,\cK)$ be an adversarial
frame. Agent $i$ maintains $f$-secrecy in $F$ if and only if, for every
interpretation $\pi$, if $\phi$ is $f$-local and nontrivial in
$M=(S,\cK,\pi)$ then $M\sat\neg K\phi$.
\end{theorem}

Of course, the characterization of confidentiality above is extremely
strong. 
A more realistic form of confidentiality should allow some form of
information to leak. Timing information, for example, might fit in
this category. In general, given any state, the adversary should be
able to rule out states for the high-level agents that are in the
distant past, or the distant future. However, formalizing the fact
that this kind of information flow is allowed is difficult in
practice. It is not obvious how to distinguish allowed timing
information flow from attacks that rely on \emph{timing channels}
\cite{r:wray91}.

Similarly, there are cases where we must permit the 
declassification of some data in order for the system to be able to
perform useful work. The classical
example used in the literature is a password-checking program: a
program that prompts the user for a password, and logs him or her in
if the password is correct. Assume the password is a high-level
piece of data. If an adversary tries to login with password $p$ and
fails, he has gained information about the true password, namely, that
it is not $p$. Most work in information flow in the past few years has
aimed at understanding this notion of declassification of data
\cite{r:pottier00,r:zdancewic01,r:chong04}. 

Finally, there are many ways in which the above definitions are
insufficiently precise. For one, they do not take the likelihood of
states into account. Assume that the adversary initially believes that
all the states of the agents are equally likely. If after some
interaction with the system, the adversary still believes that all the
high-level states are possible, but one is overwhelmingly more likely
than the others, then one could easily argue that there there has been
information leakage, although the above definitions do not capture
it. Handling these kind of flows requires more quantitative forms of
information flow properties \cite{r:gray98,r:halpern02a}.

\section*{Confidentiality and Symbolic Cryptographic Protocol
Analysis}

One thing that the framework of the last section did not take
into account is the use of 
\emph{cryptography} to hide data from the adversary. Defining
confidentiality in the presence of cryptography is more
challenging. This form of confidentiality is sometimes studied in the
literature when considering cryptographic protocols, that is,
protocols between agents that aim at exchanging messages with some
security guarantees, such as ensuring that the messages remain
confidential, or that the origin of the messages is authenticated. In
the information-flow setting, we were interested in reasoning about
what the adversary could infer about the local state of the high
agents. In this section, however, the adversary is allowed to
intercept messages, as well as forward and inject new messages into
the system. We are interested in reasoning about what the adversary
can infer about the messages he has intercepted, despite them being
perhaps encrypted. This will be reflected in the language used to
capture the confidentiality specification: to capture information
flow, the formulas involved in the specification are those whose truth
depends on the local state of the other agents; for cryptographic
protocol analysis, as we shall see, the formulas involved in the
specification are those whose truth depends on the messages
intercepted by the adversary.

There are a number of notions of confidentiality that have been
studied in the cryptographic protocol analysis literature. A common one is
based on the intuition that the adversary is not able to distinguish
between states where the agents exchange message $m$ and states
where the agents exchange some other message $m'$, for all messages $m$
and $m'$. This is the approach taken, for instance, in the spi
calculus of Abadi and Gordon
\citeyear{r:gordon99}. Phrasing it this way brings us halfway to the
framework of the last section; however, we need to take into account
that the adversary should not be able to distinguish encrypted
messages for which he does not have the corresponding decryption
key. This requires a formalization of what the adversary can do to
messages. The view we take in this section is that an adversary can do
anything short of attempting to crack encrypted messages. Thus, we
treat the particular encryption scheme used by the agents as
perfect. (We weaken this assumption in the next section.) Such an
adversary was first formalized by Dolev and Yao
\citeyear{r:dolev83}. 
Roughly speaking, a Dolev-Yao adversary can compose messages, replay
them, or decrypt them if he knows the right keys. We first define a
symbolic representation for messages, where we write $(m_1,m_2)$ for
the pairing (or concatenation) of $m_1$ and $m_2$, and $\encr{m}{k}$
for the encryption of $m$ with $k$. We write $\inv{k}$ for the inverse
key of $k$, that is, the key used to decrypt messages encrypted with
$k$. We then define a relation $\derives$, where $H\derives m$ is
interpreted as saying that the adversary can infer message $m$ from a
set $H$ of messages.  (Intuitively, $H$ is the set of messages he has
intercepted). This relation is defined using the following inference rules: 
\begin{gather*}
\Rule{m\in H}{H\derives m} \quad \Rule{H\derives\encr{m}{k}
\quad H\derives \inv{k}}{H\derives m} \quad
\Rule{H\derives (m_1, m_2)}{H\derives m_1} \quad
\Rule{H\derives (m_1, m_2)}{H\derives m_2.}
\end{gather*}
Thus, for instance, if an adversary intercepts the
messages $\encr{m}{k_1}$, $\encr{\inv{k_1}}{k_2}$, and
$\inv{k_2}$, he can derive $m$ using these inference rules, since
\[\{\encr{m}{k_1},\encr{\inv{k_1}}{k_2},\inv{k_2}\}\derives 
m.\]
(The use of a symbolic representation for messages is the source of
the name ``symbolic cryptographic protocol analysis'' given to this
style of protocol analysis.)

We can define an indistinguishability relation by following the
formalization of Abadi and Rogaway \citeyear{r:abadi02a}. (Similar
ideas appear in Abadi and Tuttle \citeyear{r:abadi91} and Syverson and
van Oorschot \citeyear{r:syverson94}.) Intuitively, the adversary cannot
distinguish two states if his local state is the same at both states,
except that we identify encrypted messages for which he does not have
the key, to capture the intuition that he cannot distinguish encrypted
messages. For simplicity, assume that the local states $s_\adv$ of the
adversary simply consist of sets of messages (intuitively, the
messages he has intercepted, along with any initial messages, such as
public keys.)

\newcommand{\floor}[1]{\lfloor #1 \rfloor}
\newcommand{\Keys}{\mathit{Keys}}

Given a message $m$ and a set of keys $K$, let $\floor{m}_K$ be
the result of replacing every indecipherable message in $m$ by 
$\wbox$. Formally, define
\begin{align*}
\floor{p}_K & = p\\
\floor{k}_K & = k\\
\floor{(m_1,m_2)}_K & = (\floor{m_1}_K,\floor{m_2}_K)\\
\floor{\encr{m}{k}}_K & = \begin{cases}
  \encr{\floor{m}_K}{k} & \text{if $\inv{k}\in K$}\\
  \squareforqed & \text{otherwise.}
			\end{cases}
\end{align*}
It is easy to check that $m'$ is a submessage of $\floor{m}_K$ if and only if $K\cup\{m\}\derives
m'$. We extend $\floor{-}$ to sets of messages $H$ by taking
$\floor{H}_K = \{\floor{m}_K \mid m\in H\}$. Define $\Keys(H)=\{k\mid
H\derives k\}$. Finally, define the indistinguishability relation
$\cK^\dy$ by taking $(s,s')\in\cK^\dy$ if and only if 
$\floor{s_\adv}_{K}=\floor{s'_\adv}_{K'}$, where $K=\Keys(s_\adv)$ and
$K'=\Keys(s'_\adv)$.  In other
words, the adversary cannot distinguish two states in which he has
intercepted different messages, where the only difference between
these messages occurs in the content of encrypted messages for which
he does not have the decryption key.

We restrict our attention to \emph{message-transmission protocols},
protocols in which the goal is for agent $1$ to send a message to
agent $2$ in a confidential way. We assume that the adversary can
intercept messages from the network, and can also forward and inject
messages into the network. We can associate with a protocol $P$ a set
$S_P$ of global states corresponding to the states that the protocol
goes through upon execution (including states that result from the
adversary intercepting, forwarding, or injecting messages). We assume
that the global states in $S_P$ include states for all the possible
messages that could be sent.  
If $\cM$ is the set of all messages that could be sent, and
$m\in\cM$, let $G(m)\subseteq S_P$ be the set of global states where
agent $1$ sends message $m$ to agent $2$. We say a
message-transmission protocol $P$ \emph{preserves message secrecy} if
for all global states $s\in S_P$ and all messages $m\in\cM$,
\[\cK^\dy(s)\cap G(m)\ne\emptyset.\]
In other words, 
every local state of the adversary is compatible with agent $1$ having
sent any possible message $m$. 

Can we capture this syntactically? Let $\Phi_0$ be a primitive
vocabulary. Say $\phi$ \emph{depends only on the message exchanged} by the
protocol if $(M,s)\sat\phi$ if and only if $(M,s')\sat\phi$, whenever
the same message is exchanged in both states $s$ and $s'$. The
following result can be proved using techniques similar to those used
to prove Theorem~\ref{t:kevin}.

\begin{theorem} 
A message-transmission protocol $P$ preserves message secrecy if and
only if, for every interpretation $\pi$, if $\phi$ depends only on the
message exchanged by $P$ and is nontrivial in $M=(S_P,\cK^\dy,\pi)$
then $M\sat\neg K\phi$.
\end{theorem}

\newcommand{\recv}{\mathit{recv}}
\newcommand{\key}{\mathit{key}}

An alternative approach, sometimes used in the literature, leads to a
specification which is easier to enforce. This approach uses the
$\derives$ relation directly in the specification. This
specification essentially says that the adversary cannot derive the
content of the message being exchanged. (This is the approach taken,
for instance, by Casper \cite{r:lowe98}, a protocol analysis tool
based on the CSP language \cite{r:hoare85}.)  Say a
message-transmission protocol $P$ \emph{preserves message DY-secrecy} if, at
every state $s\in S_P$ where the message 
exchanged is $m$, 
\[s_\adv\nderives m.\]
This specification does not require an indistinguishability relation
for the adversary, and this suggests that it can be captured by a
specification that does not use knowledge. Indeed, the specification
can be captured rather simply if we use the right language. As opposed
to the way we have been specifying things until now, this time, we
fix a particular vocabulary and a particular interpretation
$\pi_0$.  Let $\mathit{has}(m)$ be a fixed class of primitive propositions, one
per message $m$, with $\mathit{has}(m)\in\pi_0(s)$ if and only if $s_\adv\derives m$. Let
$\mathit{exchanged}(m)$ be a fixed class of primitive propositions, one per
message $m$, with $\mathit{exchanged}(m)\in\pi_0(s)$ if and only if $m$ is the message
exchanged by the protocol at state $s$. The following result follows
immediately from the definition of message DY-secrecy.
\begin{theorem}
A message-transmission protocol $P$ preserves message DY-secrecy if and only if,
for the model $M_0=(S_P,\cK,\pi_0)$ and all messages $m$, $M_0\sat
\mathit{exchanged}(m)\rimp \neg\mathit{has}(m)$. 
\end{theorem}
This specification does not use knowledge, and uses a particular model
with a fixed interpretation. In fact, it can be seen as a form of
\emph{safety property}, following the classification of properties due
to Alpern and Schneider \citeyear{r:alpern85}. Roughly speaking, a
safety property can be checked independently at all the points of the
system; the truth or falsehood of a formula at a point does not depend
on the other points of the system. This generally leads to efficient
procedures for checking the specification. It is possible to refine
the approach by considering more general ways for the adversary to
derive messages, and to formally relate the results to specifications
based on knowledge \cite{r:halpern02e}.

\section*{Confidentiality and Cryptography}

In the last section, the framework let us capture confidentiality in
cryptographic protocols, under the assumption that the encryption was
perfect; we did not allow the adversary to extract any information
from an encrypted message for which he did not have the decryption
key. Of course, in reality, encryption schemes are not perfect, and
they can possibly leak information about the message being
encrypted. In this section, we examine how we can capture the
confidentiality of encryption schemes.

Cryptography studies, among others, the properties of encryption
schemes. Modern cryptography is motivated by two basic tenets. First,
encryption schemes are concrete mathematical systems that act on
strings (often taken to be bit strings). This view leads naturally to
finer confidentiality properties than simply showing that the
adversary cannot recover the message being encrypted. Rather,
confidentiality should mean that the adversary cannot derive
\emph{any} information about the message being encrypted, including,
say, that the first bit of the message is a 1. The second tenet is
that we do not impose any restriction on the computations that the
adversary can perform on encrypted messages, aside from the fact that they
must be \emph{feasible} computations. Generally, the class of feasible
computations is the class of \emph{probabilistic polynomial time
algorithms} \cite{r:motwani95}. The definition of a probabilistic
polynomial time algorithm is asymptotic: the running time of the
algorithm is polynomial in the length of the input. Working with such
a definition of feasibility is simplified by taking the encryption
scheme itself to be defined asymptotically, where the parameterization
is given by a \emph{security parameter}. Intuitively, the larger the
security parameter, the harder it is for an adversary to get
information about encrypted messages.

The basic definition of confidentiality for an encryption scheme is
that the adversary learns nothing about the content of an encrypted
message (except possibly, for technical reasons, information about the
length of the plaintext). The definitions we use are essentially due
to Goldwasser and Micali \citeyear{r:goldwasser84}, but simplified
following Goldreich \citeyear{r:goldreich98}. In particular, we assume
a encryption scheme where the same key is used to encrypt and decrypt
messages, and where the encryption is probabilistic: encryption with a
given key yields a probability distribution over encrypted messages.
We take $E(x)$ to be the distribution of encryptions of $x$, when the
key is selected at random. Moreover, we assume that for a security
parameter $\eta$, the keys have length $\eta$, and the scheme is used
to encrypt messages of length $\eta^2$. Thus, we can simply take the
security parameter to be the length of the keys. These restrictions,
and the following definitions, are fairly technical, and I will refer
to Goldreich \citeyear{r:goldreich98,r:goldreich01} for intuitions and
more in-depth discussions.

The definition we use is that of \emph{indistinguishability of
encryptions}, which says that an adversary cannot distinguish, based
on probabilistic polynomial time tests, whether two messages encrypted
with a random key are the same message or not, even when provided with
essentially arbitrary a priori information. Formally, let $A$ be a
feasible algorithm (which we assume returns $0$ or $1$). We say a
sequence $(x_\eta,y_\eta,z_\eta)_\eta$, where
$|x_\eta|=|y_\eta|=\eta^2$ and $|z_\eta|$ is polynomial in $\eta$, is
\emph{$A$-indistinguishable} if
\[ 1-\Pr\left[ A(E(x_\eta),z_\eta) = A(E(y_\eta),z_\eta)\right]\] is a
negligible function of $\eta$, where $f(\eta)$ is \emph{negligible} in
$\eta$ if for all polynomials $p$, $f(\eta)\le 1/p(\eta)$ for all
$\eta$. In other words, two sequences are $A$-indistinguishable if the
adversary cannot really distinguish, based on the output of $A$,
whether an encrypted message is an encryption of $x_\eta$ or of
$y_\eta$, even when provided with arbitrary information $z_\eta$. (For
instance, $z_\eta$ could be the pair $(x_\eta,y_\eta)$, meaning that
even when the adversary knows that the encrypted message is the
encryption of either $x_\eta$ or $y_\eta$, he cannot tell which is the
actual message that was encrypted.) Note that we do not require the
probabilities to be equal, but that the difference should not be 
noticeable by a polynomially-bounded observer.

An encryption scheme is \emph{semantically secure} if, for all feasible
algorithms $A$, all sequences $(x_\eta,y_\eta,z_\eta)_\eta$, where
$|x_\eta|=|y_\eta|=\eta^2$ and $|z_\eta|$ is polynomial in $\eta$, are
$A$-indistinguishable.\footnote{Strictly speaking, this is the
definition of an encryption scheme having indistinguishability of
encryptions, which can be shown to be equivalent to the traditional
definition of semantic security.} One of the many achievements of
modern cryptography is to show that there are encryption schemes that are
semantically secure, assuming the existence of mathematical entities
such as one-way functions \cite{r:goldreich01}.

We can translate semantic security of an encryption scheme $C$ into
properties in an adversarial frame, where the states of the adversary
are sequences of messages (indexed by the security parameter), along
with some initial information. Formally, a
local state for the adversary is a pair
$((x_\eta)_\eta,(z_\eta)_\eta)$, where $(x_\eta)_\eta$ is the sequence
of messages to be encrypted, and $(z_\eta)_\eta$ is the sequence of a
priori information.. Let $S_C$ be the set of all states where the
adversary has such a local state.  Note that $S_C$ does not directly
model a particular protocol between agents; we are interested in
modeling properties of an encryption scheme, not a protocol. To get an
adversarial frame, we define an indistinguishability relation
$\cK^\crypt$ as follows: take $(s,s')\in\cK^\crypt$ if and only if
$(x_\eta,y_\eta,z_\eta)_\eta$ is $A$-indistinguishable for every
feasible algorithm $A$, where $s_\adv=((x_\eta)_\eta,(z_\eta)_\eta)$
and $s'_\adv=((y_\eta)_\eta,(z_\eta)_\eta)$.

We can, up to a point, capture semantic security of the encryption
scheme using a knowledge specification. 
Let $\Phi_0$ be a primitive vocabulary. Say $\phi$ \emph{depends only on
messages but not their length} in $M$ when the following properties hold:
\begin{enumerate}
\item[(1)]
if $s_\adv=s'_\adv$, then $(M,s)\sat\phi$ if and only if
$(M,s')\sat\phi$; 
\item[(2)]
there exists $s,s'$ with $s_\adv=((x_\eta)_\eta,(z_\eta)_\eta)$,
$s'_\adv=((y_\eta)_\eta,(z'_\eta)_\eta)$, and $|x_\eta|=|y_\eta|$ for
all $\eta$, such that $(M,s)\sat\phi$ and
$(M,s')\sat\neg\phi$. 
\end{enumerate}
The following result follows almost immediately from the definition of 
semantic security.

\begin{theorem}
If an encryption scheme $C$ is semantically secure, then, for every
interpretations $\pi$, if $\phi$ depends only on
messages but not on their length and is nontrivial in
$M=(S_C,\cK^\crypt,\pi)$ then $M\sat\neg K\phi$.
\end{theorem}

This formalizes one intuition behind semantic security, namely that it
ensures the adversary cannot derive any (nontrivial) knowledge about
the content of encrypted messages, except perhaps their length. It is
not clear how to get the other direction of the implication without
making stronger assumptions on the language or the models. 

This result is unsatisfying compared to the results of the previous
section as far as it concerns reasoning about protocols. In
particular, the states of the models are more ``artificial'', and do
not correspond directly to states that arise during the execution of
a protocol. A more interesting result would be to characterize the
knowledge of an adversary in the context of message-transmission
protocols implemented using an encryption scheme with a property such
as semantic security. This is an active research area. Some results
have been obtained using techniques from programming languages
\cite{r:lincoln98,r:abadi02a}, and logical techniques have been
brought to bear on the question \cite{r:impagliazzo03}, but no
connection to knowledge has yet been established, as far as I know. 

\paragraph{Acknowledgments.} Thanks to Steve Chong, Joe Halpern, Kevin
O'Neill, Sabina Petride, and Vicky Weissman for comments.

\bibliographystyle{chicagor}

\end{document}